\documentclass[english]{article}
\usepackage[T1]{fontenc}
\usepackage[latin9]{inputenc}
\usepackage[letterpaper]{geometry}
\usepackage{color}
\usepackage{array}
\usepackage{float}
\usepackage{amsmath}
\usepackage{graphicx}
\usepackage{amssymb}

\makeatletter

\providecommand{\tabularnewline}{\\}
\floatstyle{ruled}
\newfloat{algorithm}{tbp}{loa}
\floatname{algorithm}{Algorithm}

\usepackage{times}
\usepackage{mathptmx}

\makeatother

\usepackage{babel}

\begin{document}

\title{Probabilistic Models over Ordered Partitions with Application in
Learning to Rank}

\author{Tran The Truyen, Dinh Q. Phung and Svetha Venkatesh\\
Department of Computing, Curtin University\\
GPO Box U1987, Perth, Western Australia 6845, Australia\\
\{t.tran2,d.phung,s.venkatesh\}@curtin.edu.au\\
\\
Technical Report}
\maketitle
\begin{abstract}
This paper addresses the general problem of modelling and learning
rank data with ties. We propose a probabilistic generative model,
that models the process as permutations over partitions. This results
in super-exponential combinatorial state space with unknown numbers
of partitions and unknown ordering among them. We approach the problem
from the discrete choice theory, where subsets are chosen in a stagewise
manner, reducing the state space per each stage significantly. Further,
we show that with suitable parameterisation, we can still learn the
models in linear time. We evaluate the proposed models on the problem
of learning to rank with the data from the recently held Yahoo! challenge,
and demonstrate that the models are competitive against well-known
rivals.

\end{abstract}
\global\long\def\objects{\mathit{X}}

\global\long\def\bpi{\boldsymbol{\pi}}

\global\long\def\bsigma{\boldsymbol{\sigma}}

\global\long\def\rating{r}

\global\long\def\Stirling{S}

\global\long\def\Fubini{\text{Fubini}}

\global\long\def\logll{\mathcal{L}}

\global\long\def\model{\text{PMOP}}

\global\long\def\remainset{R}

\section{Introduction}

Ranking appears to be natural to humans as we often express preference
over things. Consequently, rank data has been widely studied in statistical
sciences (e.g. see \cite{marden1995analyzing} for a comprehensive
survey). More recently, the intersection between machine learning
and information retrieval has resulted in a fruitful sub-area called
learning to rank (e.g. see \cite{liu2009learning} for a recent review),
where the goal is to learn rank functions that can accurately order
objects from retrieval systems. Broadly speaking, a rank is a type
of permutation, where the ordering of objects has some meaningful
interpretation - e.g. the rank of student performance in a class.
Although we would like to obtain a complete ordering over a set of
objects, often this is possible only in small sets. In larger sets,
it is more natural to rate an object from a rating scale, and the
result is that many objects may have the same rating. Such phenomena
is common in large sets such as movies, books or web-pages wherein
many objects may have \emph{tied ratings}.

This paper focuses on the modelling and learning rank data with ties.
Previous work often involves paired comparisons (e.g. see \cite{davidson1970extending}\cite{glenn1960ties}\cite{rao1967ties}),
ignoring simultaneous interactions among objects. Such interactions
can be strong - in the case of learning to rank, objects are often
returned from a query, and thus clearly related to the query and to
each other. We take an alternative approach by modelling objects with
the same tie as a partition, translating the problem into ranking
or ordering these partitions. This problem transformation results
in a combinatorial problem- set partitioning with unknown numbers
of subsets with unknown order amongst them. For a given number of
partitions, the order amongst them is a permutation of the partitions
being considered, wherein each partition has objects of the same rank.
A generative view of the problem can then be as follows: Choose the
first partition with elements of rank $1$, then choose the next partition
from the remaining objects with elements ranked $2$ and so on. The
number of partitions then does not have to be specified in advance,
and can be treated as a random variable. The joint distribution for
each ordered partition can then be composed using a variant of the
Plackett-Luce model \cite{luce1959individual}\cite{plackett1975analysis},
substituting \emph{object }potentials by the \emph{partition }potential.
We propose two choices for these potential functions: First, we consider
the potential of each partition to be the normalised sum of individual
object potentials in that partition, leading to a simple normalisation
factor in the estimation of the joint distribution. Second, we propose
a MCMC based parameter estimation for the general choice of potential
functions. We specify this model as the Probabilistic Model over Ordered
Partitions. Demonstrating its application to the learning to rank
problem, we use the dataset from the recently held Yahoo! challenge
\cite{yahoo_challenge10}. Besides the regular first-order features,
we study second-order features constructed as the Cartesian product
over the feature set. We show that our results both in terms of predictive
performance and training time are competitive with other well-known
methods such as RankNet \cite{burges2005learning}, Ranking SVM \cite{joachims2002optimizing}
and ListMLE \cite{xia2008listwise}. With the choice of our proposed
simple potential function, we get the added advantage of lower computational
cost as it is linear in the query size compared to quadratic complexity
for the pairwise methods.

Our main contributions are the construction of a probabilistic model
over ordered partitions and associated inference and learning techniques.
The complexity of this problem is super-exponential with respect to
number of objects ($N$) because both the number of partitions and
their order are unknown - it grows exponentially as $N!/(2\left(\ln2\right)^{N+1})$
\cite[pp. 396--397]{muresan2008concrete}. Our contribution is to
overcome this computational complexity through the choice of suitable
potential functions, yielding learning algorithms with linear complexity,
thus making the algorithm deployable in real settings. The novelty
lies in the rigorous examination of probabilistic models over ordered
partitions, extending earlier work in discrete choice theory \cite{fligner1988multistage}\cite{luce1959individual}\cite{plackett1975analysis}.
The significance of the model is its potential for use in many applications.
One example is the learning to rank with ties problem and is studies
in this paper. Further, the model opens new potential applications
for example, novel types of clustering, in which the clusters are
automatically ordered.

\section{Background \label{sec:Background}}

In this section, we review some background in rank modelling and learning
to rank which are related to our work.

\paragraph{Rank models.}

Probabilistic models of permutation in general and of rank in particular
have been widely analysed in statistical sciences (e.g. \cite{marden1995analyzing}
for a comprehensive survey). Since the number of all possible permutations
over $N$ objects is $N!$, multinomial models are only computationally
feasible for small $N$ (e.g. $N\le10$). One approach to avoid this
state space explosion is to deal directly with the data space, i.e.
based on the distance between two ranks. The assumption is that there
exists a \emph{modal} ranking over all objects, and what we observe
are ranks randomly distributed around the mode. The most well-know
model is perhaps the Mallows \cite{mallows1957non}, where the probability
of a rank decreases exponentially with the distance from the mode.
Depending on the distance measures, the model may differ; and the
popular distance measures include those by Kendall and Spearman. The
problem with this approach is that it is hard to handle the cases
of multiple modes, with ties and incomplete ranking.

Another line of reasoning is largely associated with the discrete
choice theory (e.g. see \cite{luce1959individual}), which assumes
that each object has an intrinsic worth which is the basis for the
ordering between them. For example, Bradley and Terry \cite{bradley1952rank}
assumed that the probability of object preference is proportional
to its worth, resulting in the logistic style distribution for pairwise
comparison. Subsequently, Luce \cite{luce1959individual} and Plackett
\cite{plackett1975analysis} extended this model to multiple objects.
More precisely, for a set of $N$ objects denoted by $\{x_{1},x_{2},...,x_{N}\}$
the probability of ordering $x_{1}\succ x_{2}\succ...\succ x_{N}$
is defined as

\[
P(x_{1}\succ x_{2}\succ...\succ x_{N})=\prod_{i=1}^{N}\frac{\phi(x_{i})}{\sum_{j=i}^{N}\phi(x_{j})}\]
where $x_{i}\succ x_{j}$ denotes the preference of object $x_{i}$
over $x_{j}$, and $\phi(x_{i})\in\mathbb{R}$ is the worth of the
object $x_{i}$. The idea is that, we proceed in selecting objects
in a stagewise manner: Choose the first object among $N$ objects
with probability of $\phi(x_{1})/\sum_{j=1}^{N}\phi(x_{j})$, then
choose the second object among the remaining $N-1$ objects with probability
of $\phi(x_{2})/\sum_{j=2}^{N}\phi(x_{j})$ and so on until all objects
are chosen. It can be verified that the distribution is proper, that
is $P(x_{1}\succ x_{2}\succ...\succ x_{N})>0$ and the probabilities
of all possible orderings will sum to one. This paper will follow
this approach as it is easily interpretable and flexible to incorporate
ties and incomplete ranks. 

Finally, for completeness, we mention in passing the third approach,
which treats a permutation as a symmetric group and applying spectral
decomposition techniques \cite{diaconis1989generalization}\cite{huang2009fourier}.

\paragraph{Learning to rank.}

Learning-to-rank is an active topic in the intersection between machine
learning and information retrieval (e.g. see \cite{liu2009learning}
for a recent survey). The basic idea is that we can learn ranking
functions that can capture the relevance of an object (e.g. document
or image) with respect to a query. Although it appears to be an application
of rank theory, the setting and goal are inherently different from
traditional rank data in statistical sciences. Often, the pool of
all possible objects in a typical retrieval system is very large,
and often changes over time. Thus, it is not possible to enumerate
objects in the rank models. Instead, each object-query pair is associated
with a feature vector, which often describes how relevant the object
is with respect to the query. As a result, the distribution over objects
is query-specific, and these distributions share the same parameter
set. As discussed in \cite{liu2009learning}, machine learning methods
extended to ranking can be divided into:

\emph{Pointwise approach} which includes methods such as ordinal regression
\cite{chu2006gpo}\cite{cossock2008statistical}. Each query-document
pair is assigned a ordinal label, e.g. from the set $\{0,1,2,...,M\}$.
This simplifies the problem as we do not need to worry about the exponential
number of permutations. The complexity is therefore linear in the
number of query-document pairs. The drawback is that the ordering
relation between documents is not explicitly modelled.

\textit{Pairwise approach} which spans preference to binary classification
\cite{burges2005learning}\cite{freund2004eba}\cite{joachims2002optimizing}
methods, where the goal is to learn a classifier that can separate
two documents (per query). This casts the ranking problem into a standard
classification framework, wherein many algorithms are readily available,
for example, SVM \cite{joachims2002optimizing}, neural network and
logistic regression \cite{burges2005learning}, and boosting \cite{freund2004eba}.
The complexity is quadratic in number of documents per query and linear
in number of queries. Again, this approach ignores the simultaneous
interaction about objects within the same query.

\emph{Listwise approach} which models the distribution of permutations
\cite{cao2007learning}\cite{volkovs2009boltzrank}\cite{xia2008listwise}.
The ultimate goal is to model a full distribution of all permutations,
and the prediction phase outputs the most probable permutation. This
approach appears to be most natural for the ranking problem. In fact,
the methods suggested in \cite{cao2007learning}\cite{xia2008listwise}
are applications of the Plackett-Luce model.

\section{Modelling Sets with Ordered Partitions}

\begin{figure}
\begin{centering}
\includegraphics[width=0.4\linewidth]{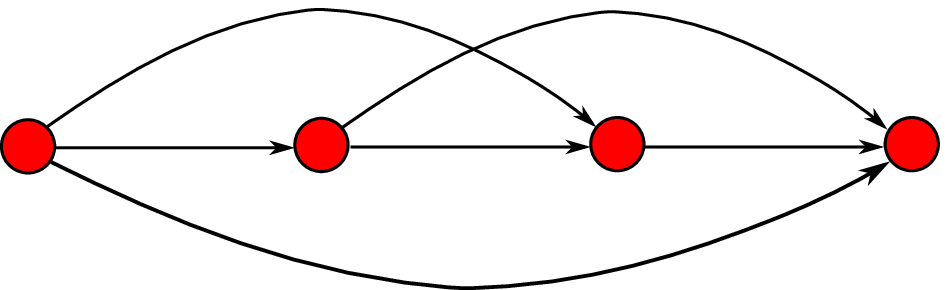}~~~~~~~~~~~~~~~~~~~~~~~~\includegraphics[width=0.4\linewidth]{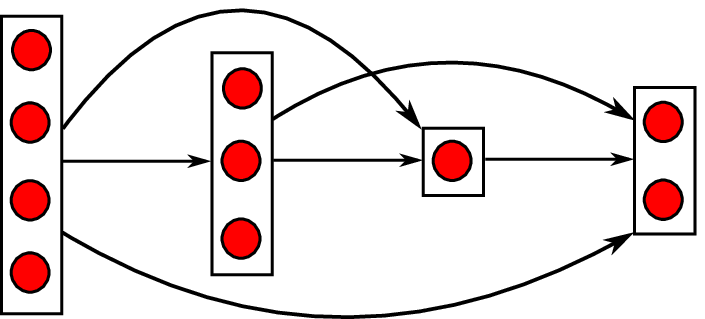}
\par\end{centering}

\caption{Complete ordering (left) versus subset ordering (right). For the subset
ordering, the bounding boxes represents the subsets of elements of
the same rank. Subset sizes are $4,3,1,2$, respectively. \label{fig:Complete-versus-subset-ordering}}

\end{figure}

\subsection{Problem Description}

Let $\objects=\left\{ x_{1},x_{2},\ldots,x_{N}\right\} $ be a collection
of $N$ objects. In a complete ranking setting, each object $x_{i}$
is further assigned with a ranking index $\pi_{i}$, resulting in
the ranked list of $\left\{ x_{\pi_{1}},x_{\pi_{2}},\ldots,x_{\pi_{N}}\right\} $
where $\bpi=\left(\pi_{1},\ldots,\pi_{N}\right)$ is a permutation
over $\left\{ 1,2,\ldots,N\right\} $. For example, $\objects$ might
be a set of documents returned by a search engine in response to a
query, and $\pi_{1}$ is the index to the first document, $\pi_{2}$
is the index to second document and so on. Ideally $\bpi$ should
contain ordering information for all returned documents; however,
this task is not always possible for any non-trivial size $N$ due
to the labor cost involved%
\footnote{We are aware that clickthrough data can help to obtain a complete
ordering, but the data may be noisy.%
}. Instead, in many situations, during training a document is \emph{rated}%
\footnote{We caution the confusion between `rating' and `ranking' here. Ranking
is the process of sorting a set of objects in an increasing or decreasing
order, whereas in `rating' each object is given with a value indicating
its preference.%
} to indicate the its degree of relevance for the query. This creates
a scenario where more than one document will be assigned to the same
rating -- a situation known as `\emph{ties}' in learning-to-rank.
When we enumerate over each object $x_{i}$ and putting those with
the same rating together, the set of $N$ objects $\objects$ can
now be viewed as being divided into $K$ partitions with each partition
is assigned with a number to indicate the its unique rank $k\in\{1,2,..,K\}$.
The ranks are obtained by sorting ratings associated with each partition
in the decreasing order. Our essential contribution in this section
is a probabilistic model over this set of partitions, learning its
parameter from data, and performing inference. 

Consider a more generic setting in which we know that objects will
be rated against an ordinal value from $1$ to $K$ but do not know
individual ratings. This means that we have to consider all possible
ways to split the set $\objects$ into exactly $K$ partitions, and
then \emph{rank} those partitions from $1$ to $K$ wherein the $k$th
partition contains all objects rated with the same value $k$. This
is the first rough description of \emph{state space} for our model.
Formally, for a given $K$ and the order among the partitions $\bsigma$,
we write the set $\objects=\left\{ x_{1},\ldots,x_{N}\right\} $ as
a union of $K$ partitions \begin{align}
\objects= & \cup_{j=1}^{K}\objects_{\sigma_{j}}\label{eq:partitions}\end{align}
where $\bsigma=(\sigma_{1},\ldots,\sigma_{K})$ is a permutation over
$\left\{ 1,2,..,K\right\} $ and each partition $\objects_{k}$ is
a non-empty subset of objects with the same rating $k$. These partitions
are pairwise disjoint and having cardinality range from $1$ to $N$.
It is easy to see that when $K=N$, each $\objects_{k}$ is a singleton,
$\bsigma$ is now a complete permutation over $\left\{ 1,\ldots,N\right\} $
and the problem reduces exactly to the complete ranking setting mentioned
earlier. To get an idea of the state space, it is not hard to see
that there are $\left|\begin{array}{c}
N\\
K\end{array}\right|K!$ ways to partition and order $\objects$ where $\left|\begin{array}{c}
N\\
K\end{array}\right|$ is the number of possible ways to divide a set of $N$ objects into
$K$ partitions, otherwise known as \emph{Stirling numbers of second
kind} \cite[p. 105]{van1992course}. If we consider all the possible
values of $K$, the size of our state space is \begin{align}
\sum_{k=1}^{N}\left|\begin{array}{c}
N\\
k\end{array}\right|k! & =\Fubini\left(N\right)=\sum_{j=1}^{\infty}\frac{j^{N}}{2^{j+1}}\label{eq:Fubini_number}\end{align}
which is also known in combinatorics as the Fubini's number \cite[pp. 396--397]{muresan2008concrete}.
This is a super-exponential growth number. For instance, $\Fubini\left(1\right)=1$,
$\Fubini\left(3\right)=13$, $\Fubini\left(5\right)=541$ and $\Fubini\left(10\right)=102,247,563$.
Its asymptotic behaviour can also be shown \cite[pp. 396--397]{muresan2008concrete}
to approach $N!/(2\left(\ln2\right)^{N+1})$ as $N\rightarrow\infty$
where we note that $\ln\left(2\right)<1$, and thus it grows much
faster than $N!$. Clearly, for unknown $K$ this presents a very
challenging problem. In this paper, we shall present an efficient
and a generic approach to tackle this state-space explosion.

\subsection{Probabilistic Model over Ordered Partitions}

Return to our problem, our task now to model a distribution over the
ordered partitioning of set $\objects$ into $K$ partitions and the
ordering $\bsigma=(\sigma_{1},\ldots,\sigma_{K})$ among $K$ partitions
given in Eq (\ref{eq:partitions}): \begin{align}
p\left(\objects\right) & =p\left(\objects_{\sigma_{1}},\ldots,\objects_{\sigma_{K}}\right)\label{eq:set_prob}\end{align}
A two-stage view has been given thus far: first $\objects$ is partitioned
in any arbitrary way so long as it creates $K$ partitions and then
these partitions are ranked, result in a ranking index vector $\bsigma$.
This description is generic and one can proceed in different ways
to further characterise Eq (\ref{eq:set_prob}). We present here a
generative, multistage view to this same problem so that it lends
naturally to the specification of the distribution in Eq (\ref{eq:prob_factorize}):
First, we construct a subset $\objects_{1}$ from $\objects$ by collecting
all objects which (supposedly) have the largest ratings. If there
are more elements in the the remainder set $\left\{ \objects\setminus\objects_{1}\right\} $
to be selected, we construct a subset $\objects_{2}$ from $\left\{ \objects\setminus\objects_{1}\right\} $
whose elements have the second largest ratings. This process continues
until there is no more object to be selected.%
\footnote{This process resembles the generative process of Plackett-Luce discrete
choice model \cite{luce1959individual}\cite{plackett1975analysis},
except we apply on partitions rather than single element. It clear
from here that Plackett-Luce model is a special case of ours wherein
each partition $\objects_{k}$ reduces to a singleton. %
} An advantage of this view is that the resulting total number of partitions
$K_{\sigma}$ is automatically generated, no need to be specified
in advance and can be treated as a random variable. If our data truly
contains $K$ partitions then $K_{\sigma}$ should be equal to $K$.
Using the chain rule, we write the joint distribution over $K_{\sigma}$
ranked partitions as\begin{align}
p\left(\objects_{1},\ldots,\objects_{K_{\sigma}}\right) & =p\left(\objects_{1}\right)\prod_{k=2}^{K_{\sigma}}p\left(\objects_{k}\mid\objects_{1},\ldots,\objects_{k-1}\right)=p_{1}\left(X_{1}\right)\prod_{k=2}^{K_{\sigma}}p_{k}\left(\objects_{k}\mid\objects_{1:k-1}\right)\label{eq:prob_factorize}\end{align}
where we have used $\objects_{1:k-1}=\left\{ \objects_{1},\ldots,\objects_{k-1}\right\} $
for brevity.

\subsection{Parameterisation, Learning and Inference \label{sub:local-MCMC}}

It remains to specify the local distribution $P(\objects_{k}\mid\objects_{1:k-1})$.
Let us first consider what choices do we have after the first $\left(k-1\right)$
partitions have been selected. It is clear that we can select any
objects from the remainder set $\left\{ \objects\setminus\objects_{1:k-1}\right\} $
for our next partition $k$th. If we denote this remainder set by
$\remainset_{k}=\left\{ \objects\setminus\objects_{1:k-1}\right\} $
and $N_{k}=\left|\remainset_{k}\right|$ is the number of remaining
objects, then our next partition$\objects_{k}$ is a subset of $\remainset_{k}$;
furthermore, there is precisely $\left(2^{N_{k}}-1\right)$ such non-empty
subsets. Using the notation $2^{\remainset_{k}}$ to denote the power
set of the set $\remainset_{k}$, i.e, $2^{\remainset_{k}}$ contains
all possible non-empty subsets%
\footnote{The usual understanding would also contain the empty set, but we exclude
it in this paper.%
} of $R$, we are ready to specify each local conditional distribution
in Eq (\ref{eq:prob_factorize}) as:

\begin{equation}
p_{k}\left(\objects_{k}\mid\objects_{1:k-1}\right)=\frac{\Phi_{k}\left(\objects_{k}\right)}{\underset{S\in2^{\remainset_{k}}}{\Sigma}\Phi_{k}(S)}\label{eq:local-prob}\end{equation}
where $\Phi_{k}\left(S\right)>0$ is an order-invariant%
\footnote{i.e., the function value does not depend on the order of elements
within the partition.%
} set function defined over a set or partition $S$, and the summation
in the denominator clearly makes the definition in Eq(\ref{eq:local-prob})
a proper distribution. The set function $\Phi_{k}\left(\cdot\right)$
can also be interpreted as the potential function in standard probabilistic
graphical models literature.

Although the state space $2^{\remainset_{k}}$ for this local conditional
distribution is significantly smaller than the space of all possible
ordered partitions of $N$ objects, it is still exponential as we
have shown earlier to be $2^{N_{k}}-1$. In general, directly computing
the normalising term is still not possible, let alone learning the
model parameters. In what follows, we will study an efficient special
case which has (sub)-quadratic complexity in learning, and a general
case with MCMC approximation. \emph{We further term our Probabilistic
Model over Ordered Partition as $\model$}.

\subsubsection{Full-Decomposition $\model$ \label{sub:Full-Decomposition}}

Under a full-decomposition setting, we assume the following local
\emph{additive} decomposition at each $k$th step: \begin{equation}
\Phi_{k}(\objects_{k})=\frac{1}{\left|\objects_{k}\right|}\sum_{x\in\objects_{k}}\phi_{k}(x)\label{eq:ff_local}\end{equation}
The normalising term $\left|\objects_{k}\right|$ is to ensure that
the probability is not monotonically increasing with number of objects
in the partition. Given this form, the local normalisation factor
represented in the denominator of Eq (\ref{eq:local-prob}) can now
efficiently represented as the sum of all weighted sums of objects.
Since each object $x$ in the remainder set $\remainset_{k}$ participates
in the \emph{same} additive manner towards the construction of the
denominator in Eq (\ref{eq:local-prob}), it must admit the following
form%
\footnote{To illustrate this intuition, suppose the remainder set is $\remainset_{k}=\left\{ a,b\right\} $,
hence its power set, excluding $\emptyset$, contains 3 subsets $\left\{ a\right\} ,\left\{ b\right\} ,\left\{ a,b\right\} $.
Under the full-decomposition assumption, the denominator in Eq (\ref{eq:local-prob})
becomes $\phi\left(r_{a}\right)+\phi\left(r_{b}\right)+\frac{1}{2}\left\{ \phi\left(r_{a}\right)+\phi\left(r_{b}\right)\right\} =(1+\frac{1}{2})\sum_{x\in\left\{ a,b\right\} }\phi\left(r_{x}\right)$.
The constant term is $C=\frac{3}{2}$ in this case.%
}:

\begin{equation}
\sum_{S\in2^{\remainset_{k}}}\Phi_{k}(S)=\sum_{S\in2^{\remainset_{k}}}\frac{1}{\left|S\right|}\sum_{x\in S}\phi_{k}\left(x\right)=C\times\sum_{x\in R_{k}}\phi_{k}(x)\label{eq:decompose}\end{equation}
where $C$ is some constant and its exact value is not essential under
a maximum likelihood parameter learning treatment (readers are referred
to Appendix~\ref{sec:Computing-constant} for the computation of
$C$). To see this, substitute Eq (\ref{eq:ff_local}) and (\ref{eq:decompose})
into Eq (\ref{eq:local-prob}):

\begin{align}
\log p\left(\objects_{k}\mid\objects_{1:k-1}\right) & =\log\frac{\Phi_{k}\left(\objects_{k}\right)}{\underset{S\in2^{\remainset_{k}}}{\Sigma}\Phi_{k}(S)}=\log\frac{1}{C\left|X_{k}\right|}\frac{\sum_{x\in\objects_{k}}\phi_{k}(x)}{\sum_{x\in R_{k}}\phi_{k}(x)}=\log\frac{\sum_{x\in\objects_{k}}\phi_{k}(x)}{\sum_{x\in R_{k}}\phi_{k}(x)}-\log C\left|\objects_{k}\right|\label{eq:local_logprob}\end{align}
Since $\log C\left|\objects_{k}\right|$ is a constant w.r.t the parameters
used to parameterise the potential functions $\phi_{k}(\cdot)$, it
does not affect the gradient of the log-likelihood. It is also clear
that maximising the likelihood given in Eq (\ref{eq:prob_factorize})
is equivalent to maximising each local log-likelihood function given
in Eq (\ref{eq:local_logprob}) for each $k$. Discarding the constant
term in Eq (\ref{eq:local_logprob}), we re-write it in this simpler
form:\begin{align}
\log p\left(\objects_{k}\mid\objects_{1:k-1}\right) & =\log\sum_{x\in\objects_{k}}g_{k}\left(x\mid\objects_{1:k-1}\right)\ \text{ where }\ g_{k}\left(x\mid\objects_{1:k-1}\right)=\frac{\phi_{k}(x)}{\sum_{x\in R_{k}}\phi_{k}(x)}\label{eq:local_logprob2}\end{align}
Depend on the specific form chosen for $\phi_{k}(x)$, maximising
log-likelihood in the form of Eq (\ref{eq:local_logprob2}) can be
carried on in most cases. Gradient-based learning this type of model
is generally takes $N^{2}$ time complexity . \emph{However, using
dynamic programming technique, we show that if the function $\phi_{k}\left(x\right)$
does not depend on its position $k$, then the gradient-based learning
complexity can be reduced to linear in $N$.}

To see how, dropping the explicit dependency of the subscript $k$
in the definition of $\phi_{k}\left(\cdot\right)$, we maintain an
auxiliary array $a_{k}=\sum_{x\in R_{k}}\phi\left(x\right)$ where
$a_{K_{\sigma}}=\sum_{x\in X_{K_{\sigma}}}\phi\left(x\right)$ and
$a_{k}=a_{k+1}+\sum_{x\in X_{k}}\phi\left(x\right)$ for $k<K_{\sigma}$.
Clearly $a_{1:K_{\sigma}}$ can be computed in $N$ time in a backward
fashion. Thus, $g_{k}\left(\cdot\right)$ in Eq (\ref{eq:local_logprob2})
can also be computed linearly via the relation $g_{k}\left(x\right)=\phi\left(x\right)/a_{k}$.
This also implies that the total log-likelihood can also computed
linearly in $N$. 

Furthermore, the gradient of log-likelihood function can also be computed
linearly in $N$. Given the likelihood function in Eq (\ref{eq:prob_factorize}),
using Eq (\ref{eq:local_logprob2}), the log-likelihood function and
its gradient, without explicit mention of the parameters, can be shown
to be%
\footnote{To be more precise, for $k=1$ we define $\objects_{1:0}$ to be $\emptyset$.%
}\begin{align}
\logll & =\log p\left(\objects_{1},\ldots,\objects_{K_{\sigma}}\right)=\sum_{k=1}^{K}\log\sum_{x\in\objects_{k}}g_{k}\left(x\mid\objects_{1:k-1}\right)=\sum_{k=1}^{K}\log\sum_{x\in\objects_{k}}\frac{\phi(x)}{a_{k}}\label{eq:logll}\\
\partial\logll & =\sum_{k}\partial\log\sum_{x\in\objects_{k}}\phi\left(x\right)-\sum_{k}\partial\log a_{k}=\sum_{k}\frac{\sum_{x\in\objects_{k}}\partial\phi\left(x\right)}{\sum_{x\in\objects_{k}}\phi\left(x\right)}-\sum_{k}\frac{1}{a_{k}}\sum_{x\in R_{k}}\partial\phi\left(x\right)\label{eq:logll_gradient}\end{align}
It is clear that the first summation over $k$ in the RHS of the last
equation takes exactly $N$ time since $\sum_{k=1}^{K}\left|X_{k}\right|=N$.
For the second summation over $k$, it is more involved because both
$k$ and $R_{k}$ can possibly range from $1$ to $N$, so direct
computation will cost at most $N(N-1)/2$ time. Similar to the case
of $a_{k}$, we now maintain an 2-D auxiliary array%
\footnote{This is 2-D because we also need to index the parameters as well as
the subsets.%
} $b_{k}=\sum_{x\in R_{k}}\partial\phi(x)$, where $b_{K_{\sigma}}=\sum_{x\in X_{K_{\sigma}}}\partial\phi\left(x\right)$
and $b_{k}=b_{k+1}+\sum_{x\in X_{k}}\partial\phi\left(x\right)$ for
$k<K_{\sigma}$. Thus, $b_{1:K_{\sigma}}$, and therefore the gradient
$\partial\logll$, can be computed in $NF$ time in a backward fashion,
where $F$ is the number of parameters.

\subsubsection{General State $\model$ and MCMC Inference \label{sub:General-State}}

In the general case without any assumption on the form of the potential
function $\Phi_{k}\left(\cdot\right)$ using only Eq (\ref{eq:local-prob})
and (\ref{eq:prob_factorize}), the log-likelihood function and its
gradient, again without explicit mention of the model parameter, are:\begin{align}
\logll & =\log p\left(\objects_{1}\right)+\sum_{k=2}^{K_{\sigma}}\log p_{k}\left(\objects_{k}\mid\objects_{1:k-1}\right)\label{eq:general_logll}\\
\partial\logll & =\sum_{k=1}^{K_{\sigma}}\partial\log\Phi_{k}\left(\objects_{k}\right)-\sum_{k=1}^{K_{\sigma}}\left\{ \sum_{S\in2^{R_{k}}}p_{k}\left(S\mid\objects_{1:k-1}\right)\partial\log\Phi_{k}\left(S\right)\right\} \label{eq:general_logll_grad}\end{align}
Clearly, both the distribution $p_{k}\left(\objects_{k}\mid\objects_{1:k-1}\right)$
and the expectation $\sum_{S\in2^{R_{k}}}p_{k}\left(S\mid\objects_{1:k-1}\right)\partial\log\Phi_{k}\left(S\right)$
are generally intractable to evaluate. In this paper, we make use
of MCMC methods to approximate $p_{k}\left(\objects_{k}\mid\objects_{1:k-1}\right)$.
There are two natural choices: the Gibbs sampling and Metropolis-Hastings
sampling. For Gibbs sampling we note that this problem can be viewed
as sampling from a random field with binary variables. Each object
is attached with binary variable whose states are either `\emph{selected}'
or `\emph{not selected}' at $k$th stage. Thus, there will be $2^{N_{k}}-1$
joint states in the random field, where we recall that $N_{k}$ is
the total number of remaining objects after $(k-1)$-th stage. The
pseudo code for Gibbs and Metropolis-Hastings routines performed at
$k$th stage is illustrated in Alg. (\ref{alg:MCMC}).

\begin{algorithm}[h]
\begin{centering}
\begin{tabular}{>{\raggedright}p{0.5\textwidth}|>{\raggedright}p{0.5\textwidth}}
\textbf{Gibbs sampling }\linebreak{}

1. Randomly choose an initial subset $X_{k}$

2. Repeat until stopping criteria met
\begin{itemize}
\item For each remaining object $x$ at stage $k$, randomly select the
object with the probability \[
\frac{\Phi_{k}(X_{k}^{+x})}{\Phi_{k}(X_{k}^{+x})+\Phi_{k}(X_{k}^{-x})}\]
where $\Phi_{k}(X_{k}^{+x})$ is the potential of the currently selected
subset $X_{k}$ if $x$ is included and $\Phi_{k}(X_{k}^{-x})$ is
when $x$ is not.
\end{itemize}
 & \textbf{Metropolis-Hastings sampling}\linebreak{}

1. Randomly choose an initial subset $\objects_{k}$

2. Repeat until stopping criteria met
\begin{itemize}
\item Randomly choose number of objects $m$, subject to $1\le m\le N_{k}$. 
\item Randomly choose $m$ distinct objects from remaining set $R_{k}=\left\{ \objects\setminus\objects_{1:k-1}\right\} $
to construct a new partition denoted by $S$
\item Set $\objects_{k}\leftarrow S$ with the probability of $\min\left\{ 1,\frac{\Phi_{k}(S)}{\Phi_{k}(\objects_{k})}\right\} $
\end{itemize}
\tabularnewline
\end{tabular}
\par\end{centering}

\caption{MCMC sampling approaches for PMOP in general case.\label{alg:MCMC}}

\end{algorithm}

Finally, we note that in practical implementation of learning, we
follow the proposal in \cite{Hinton02} wherein for each local distribution
at $k$th round we run the MCMC for \emph{only} a few steps starting
from the observed subset $\objects_{k}$. This technique is known
to produce a biased estimate, but empirical evidences have so far
indicated that the bias is small and the estimate is effective. Importantly,
it is very fast compared to full sampling.

\subsection{Learning-to-Rank with $\model$\label{sub:Learning-to-rank}}

To conclude the presentation of our proposed model for probabilistic
modelling over ordered partitions (PMOP), we present a specific application
of PMOP for the problem of leaning-to-rank. The ultimate goal after
training is that, for each query the system needs to return a list
of related objects and their \emph{ranking}.%
\footnote{We note a confusion that may arise here is that, although during training
each training query $q$ is supplied with a list of related objects
and their \emph{ratings,} during the ranking phase the system still
needs to return a ranking over the list of related objects for an
unseen query.%
} Slightly different from the standard rank setting in statistics,
the objects in learning-to-rank problem are often not indexed (e.g.
the identity of the object is not captured in any parameter). Instead,
we will assume that for each query-object pair $\left(q,x\right)$
we can extract a feature vector $x^{q}$. Model distribution specified
in this way is thus\emph{ query-specific}. As a result, we are not
interested in finding the single mode for the rank distribution over
all queries%
\footnote{This would lead to something like the \emph{static} rank over all
possible objects in the database - like those in Google's PageRank
\cite{brin1998anatomy}.%
}, but in finding the rank mode for each query. 

At the ranking phase, suppose for a unseen query $q$ a list of $\objects^{q}=\left\{ x_{1}^{q},\ldots,x_{N_{q}}^{q}\right\} $
objects related to $q$ is returned. The task is then to rank these
objects in decreasing order of relevance w.r.t $q$. Enumerating over
all possible ranking take an order of $N_{q}!$ time. Instead we would
like to establish a \emph{scoring function} $f(x^{q},w)\in\mathbb{R}$
for the query $q$ and each object $x$ returned where $w$ is now
introduced as the parameter. Sorting can then be carried out much
more efficiently in the complexity order of $N_{q}\log N_{q}$ instead
of $N_{q}!$. The function specification can be a simple a linear
combination of features $f(x^{q},w)=w^{\top}x^{q}$ or more complicated
form, such as a multilayer neural network, can be used.

In the practice of learning-to-rank, the dimensionality of feature
vector $x^{q}$ is often remains the same across all queries, and
since it is observed, we use $\model$ described before to specify
conditional model specific to $q$ over the set of returned objects
$\objects^{q}$ as follows. \begin{equation}
p\left(\objects^{q}|w\right)=p(\objects_{1}^{q},\objects_{2}^{q},...,\objects_{K_{\sigma}}^{q}\mid w)=P(\objects_{1}^{q}\mid w)\prod_{k=2}^{K_{\sigma}}p(\objects_{k}^{q}\mid\objects_{1:k-1}^{q},w)\label{eq:rank_factorization}\end{equation}
We can see that Eq (\ref{eq:rank_factorization}) has exactly the
same form of Eq (\ref{eq:prob_factorize}) specified for $\model$,
but applied instead on the query-specific set of objects $\objects^{q}$
and additional parameter $w$. During training, each query-object
pair is labelled by a relevance score, which is typically an integer
from the set $\{0,..,M\}$ where $0$ means the object is irrelevant
w.r.t the query $q$, and $M$ means the object is highly relevant%
\footnote{Note that generally $K\ne M+1$ because there may be gaps in rating
scales for a specific query.%
}. The value of $M$ is typically much smaller than $N_{q}$, thus,
the issue of \emph{ties,} described at the beginning of this section,
occur frequently. In a nutshell, for each training query $q$ and
its rated associated list of objects a PMOP is created. \emph{The
important parameterisation to note here is that the parameter $w$
is shared across all queries}; and thus, enabling ranking for unseen
query in the future.

Using the scoring function $f\left(x,w\right)$ we specify the individual
potential function $\phi\left(\cdot\right)$ in the exponential form:\begin{align*}
\phi_{k}\left(x,w\right) & =\exp\left\{ f\left(x,w\right)\right\} \end{align*}
The local potential function defined over for partition $\Phi_{k}\left(\objects_{k}^{q}\right)$
can now be explicitly constructed under full-decomposition (Subsection~\ref{sub:Full-Decomposition})
and general case (Subsection~\ref{sub:General-State}) as respectively
follows.

\begin{eqnarray}
\text{Full-decomposition: }\ \Phi_{k}\left(\objects_{k}^{q}\right) & = & \frac{1}{|\objects_{k}^{q}|}\sum_{x\in\objects_{k}^{q}}\exp\left\{ f(x,w)\right\} \label{eq:rank_full_decompose}\end{eqnarray}

\begin{equation}
\text{General case: }\ \Phi_{k}\left(\objects_{k}^{q}\right)=\exp\left\{ \frac{1}{|\objects_{k}^{q}|}\sum_{x\in\objects_{k}^{q}}f\left(x,w\right)\right\} \label{eq:rank_general}\end{equation}

The gradient of the log-likelihood function can also be computed efficiently.
For full-decomposition, it can be shown to be:\begin{align*}
\frac{\partial\log p\left(\objects_{k}^{q}\mid\objects_{1:k-1}^{q}\right)}{\partial w} & =\sum_{x\in\objects_{k}^{q}}\frac{\phi_{k}(x,w)x}{\sum_{x\in\objects_{k}^{q}}\phi_{k}(x,w)}-\sum_{x\in R_{k}^{q}}\frac{\phi_{k}(x,w)x}{\sum_{x\in R_{k}^{q}}\phi_{k}(x,w)}\end{align*}

For the general case, the gradient of the log-likelihood function
can be shown to be:\begin{align*}
\frac{\partial\log p\left(\objects_{k}^{q}\mid\objects_{1:k-1}^{q}\right)}{\partial w} & =\bar{x}_{k}^{q}-\sum_{S_{k}\in2^{R_{k}^{q}}}p\left(S_{k}\mid\objects_{1:k-1}^{q}\right)\bar{s}_{k}\end{align*}
where\begin{align*}
\bar{x}_{k}^{q} & =\frac{1}{|X_{k}^{q}|}\sum_{x\in X_{k}}x^{q}\end{align*}
The quantity $p\left(X_{k}^{q}\mid\objects_{1:k-1}^{q}\right)$ can
be interpreted as the probability that the subset $X_{k}^{q}$ is
chosen out of all possible subsets at stage $k$, and $\bar{x}_{k}$
is the centre of the chosen subset. 

The expectation $\sum_{S_{k}}P(S_{k}\mid\objects_{1:k-1}^{q})\bar{s}_{k}$
is expensive to evaluate, since there are $2^{N_{k}}-1$ possible
subsets. Thus, we resort to MCMC techniques. We follow the suggestion
in \cite{Hinton02} to start the Markov chain from the observed subset
$X_{k}$ and run for a few iterations. The parameter update is stochastic\[
w\leftarrow w+\eta\sum_{k}\left(\bar{x}_{k}^{q}-\frac{1}{n}\sum_{l=1}^{n}\bar{s}_{k}^{(l)}\right)\]
where $\bar{s}_{k}^{(l)}$is the centre of the subset sampled at iteration
$l$, and $\eta>0$ is the learning rate, and $n$ is number of samples.
Typically we choose $n$ to be small, e.g. $n=1,2,3$.






\section{Discussion \label{sec:Discussion}}

In our specific choice of the local distribution in Eq (\ref{eq:local-prob}),
we share the same idea with that of Plackett-Luce, in which the probability
of choosing the subset is proportional to the subset's worth, which
is realised by the subset potential. In fact, when we limit the subset
size to $1$, i.e. there are no ties, the proposed model reduces to
the well-known Plackett-Luce models. 

It is worth mentioning that the factorisation in Eq (\ref{eq:prob_factorize})
and the choice of local distribution in Eq (\ref{eq:local-prob})
are not unique. In fact, the chain-rule can be applied to any sequence
of choices. For example, we can factorise in a backward manner \begin{align}
p\left(\objects_{1},\ldots,\objects_{K_{\sigma}}\right) & =p_{1}\left(X_{K_{\sigma}}\right)\prod_{k=1}^{K_{\sigma}-1}p_{k}\left(\objects_{k}\mid\objects_{k+1:K_{\sigma}}\right)\label{eq:prob_factorize}\end{align}
where $\objects_{k+1:K_{\sigma}}$ is a shorthand for $\{X_{k+1},X_{k+2},...,X_{K_{\sigma}}\}$.
Interestingly, we can interpret this reverse process as \emph{subset
elimination}: First we choose to eliminate the worst subset, then
the second worst, and so on. This line of reasoning has been discussed
in \cite{fligner1988multistage} but it is limited to $1$-element
subsets. However, if we are free to choose the parameterisation of
$p_{k}\left(\objects_{k}\mid\objects_{k+1:K_{\sigma}}\right)$ as
we have done for $p_{k}\left(\objects_{k}\mid\objects_{1:k-1}\right)$
in Eq (\ref{eq:local-prob}), there are not guarantee that the forward
and backward factorisations admit the same distribution. 

Our model can be placed into the framework of probabilistic graphical
models (e.g. see \cite{Lauritzen96}\cite{Pearl88}). Recall that
in standard probabilistic graphical models, we have a set of variables,
each of which receives values from a fixed set of states. Generally,
variables and states are orthogonal concepts, and the \emph{state
space} of a variable do not explicitly depends on the states of other
variables%
\footnote{Note that, this is different from saying the states of variables are
independent.%
}. In our setting, the objects play the role of the variables, and
their memberships in the subsets are their states. However, since
there are exponentially many subsets, enumerating the state spaces
as in standard graphical models is not possible. Instead, we can consider
the ranks of the subsets in the list as the states, since the ranks
only range from $1$ to $N$. Different from the standard graphical
models, the variables and the states are not always independent, e.g.
when the subset sizes are limited to $1$, then the state assignments
of variables are mutually exclusive, since for each position, there
is only one object. Probabilistic graphical models are generally directed
(such as Bayesian networks) or undirected (such as Markov random fields),
and our $\model$ can be thought as a directed model. The undirected
setting is also of great interest, but it is beyond the scope of this
paper.

With respect to tie handling, most previous work focuses on pairwise
models. The basic idea is to assign some probability mass for the
event of ties \cite{davidson1970extending}\cite{glenn1960ties}\cite{rao1967ties}.
For instance, denote by $x_{i}\succ x_{j}$ the preference of $x_{i}$
over $x_{j}$, and by $x_{i}\approx x_{j}$ the tie between the two
objects, Rao and Kupper \cite{rao1967ties} proposed the following
models \begin{eqnarray}
P(x_{i}\succ x_{j}) & = & \frac{\phi(x_{i})}{\phi(x_{i})+\theta\phi(x_{j})}\nonumber \\
P(x_{i}\approx x_{j}) & = & \frac{(\theta^{2}-1)\phi(x_{i})\phi(x_{j})}{\left[\phi(x_{i})+\theta\phi(x_{j})\right]\left[\theta\phi(x_{i})+\phi(x_{j})\right]}\label{eq:Rao-Kupper}\end{eqnarray}
where $\theta\ge1$ is the parameter to control the contribution of
ties. When $\theta=1$, the model reduces to the standard Bradley-Terry
model \cite{bradley1952rank} . This method of ties handling is further
studied in \cite{zhou2008learning} in the context of learning to
rank. Another method is introduced in \cite{davidson1970extending},
where the probability masses are defined as\begin{eqnarray}
P(x_{i}\succ x_{j}) & = & \frac{\phi(x_{i})}{\phi(x_{i})+\phi(x_{j})+\nu\sqrt{\phi(x_{i})\phi(x_{j})}}\nonumber \\
P(x_{i}\approx x_{j}) & = & \frac{\nu\sqrt{\phi(x_{i})\phi(x_{j})}}{\phi(x_{i})+\phi(x_{j})+\nu\sqrt{\phi(x_{i})\phi(x_{j})}}\label{eq:Davidson}\end{eqnarray}
where $\nu\ge0$. The applications of these two tie-handling models
to learning to rank are detailed in Appendix~\ref{sec:Learning-the-Paired-Ties}.

For ties of multiple objects, we can create a group of objects, and
work directly on groups. For example, let $X_{i}$ and $X_{j}$ be
two sport teams, the pairwise team ordering can be defined using the
Bradley-Terry model as \[
P(X_{i}\succ X_{j})=\frac{\sum_{x\in X_{i}}\phi(x)}{\sum_{x\in X_{i}}\phi(x)+\sum_{s\in X_{j}}\phi(s)}\]
The extension of the Plackett-Luce model to multiple groups has been
discussed in \cite{huang2006generalized}. However, we should emphasize
that this setting is not the same as ours, because the partitioning
is known in advance, and the groups behave just like standard super-objects.
Our setting, on the other hand, assumes no fixed partitioning, and
the membership of the objects in a group is arbitrary.

\section{Evaluation}

\subsection{Setting}

The data is from Yahoo! learning to rank challenge \cite{yahoo_challenge10}.
This is currently the largest dataset available for research. At the
time of this writing, the data contains the groundtruth labels of
$473,134$ documents returned from $19,944$ queries. The label is
the relevance judgment from $0$ (irrelevant) to $4$ (perfectly relevant).
Features for each document-query pairs are also supplied by Yahoo!,
and there are $519$ unique features.

We split the data into two sets: the training set contains roughly
$90\%$ queries, and the test set is the remaining $10\%$. Two performance
metrics are reported: the Normalised Discounted Cumulative Gain at
position $T$ (NDCG$@T$), and the Expected Reciprocal Rank (ERR).
NDCG$@T$ metric is defined as

\begin{eqnarray*}
\mbox{NDCG}@T & =\frac{1}{\kappa(T)} & \sum_{i=1}^{T}\frac{2^{r_{i}}-1}{\log_{2}(1+i)}\end{eqnarray*}
where $r_{i}$ is the relevance judgment of the document at position
$i$, $\kappa(T)$ is a normalisation constant to make sure that the
gain is $1$ if the rank is correct. The ERR is defined as\begin{eqnarray*}
\mbox{ERR} & = & \sum_{i}\frac{1}{i}V(r_{i})\prod_{j=1}^{i-1}(1-V(r_{j}))\,\,\,\,\,\mbox{where}\,\, V(r)=\frac{2^{r}-1}{16}\end{eqnarray*}
which puts even more emphasis on the top-ranked documents.

For comparison, we implement several well-known methods, including
RankNet \cite{burges2005learning}, Ranking SVM \cite{joachims2002optimizing}
and ListMLE \cite{xia2008listwise}. The RankNet and Ranking SVM are
pairwise methods, and they differ on the choice of loss functions,
i.e. logistic loss for the RankNet and hinge loss for the Ranking
SVM%
\footnote{Strictly speaking, RankNet makes use of neural networks as the scoring
function, but the overall loss is still logistic, and for simplicity,
we use simple perceptron.%
}. Similarly, choosing quadratic loss gives us a rank regression method,
which we will call Rank Regress. From rank modelling point of view,
the RankNet is essentially the Bradley-Terry model \cite{bradley1952rank}
applied to learning to rank. Likewise, the ListMLE is essentially
the Plackett-Luce model. We also implement two variants of the Bradley-Terry
model with ties handling, one by Rao-Kupper \cite{rao1967ties} (denoted
by PairTies-RK; this also appears to be implemented in \cite{zhou2008learning}
under the functional gradient setting) and another by Davidson \cite{davidson1970extending}
(denoted by PairTies-D; and this is the first time the Davidson method
is applied to learning to rank). See Appendix~\ref{sec:Learning-the-Paired-Ties}
for implementation details.

There are three methods resulted from our framework (see description
in Section~\ref{sub:Learning-to-rank}). The first is the $\model$
with full-decomposition (denoted by $\model$-FD), the second is with
Gibbs sampling (denoted by $\model$-Gibbs), and the third is with
Metropolis-Hastings sampling (denoted by $\model$-MH). 

For those pairwise methods without ties handling, we simply ignore
the tied document pairs. For the ListMLE, we simply sort the documents
within a query by relevance scores, and those with ties are ordered
according to the sorting algorithm. All methods, except for $\model$-Gibbs/MH,
are trained using the Limited Memory Newton Method known as L-BFGS.
The L-BFGS is stopped if the relative improvement over the loss is
less than $10^{-5}$ or after $100$ iterations. As the $\model$-Gibbs/MH
are stochastic, we run the MCMC for a few steps per query, then update
the parameter using the Stochastic Gradient Ascent. The learning rate
is fixed to $0.1$, and the learning is stopped after $1,000$ iterations.

As for feature representation, we first normalised the features across
the whole training set to roughly have mean $0$ and standard deviation
$1$. We then employ both the first-order features and second-order
features (by taking the Cartesian product of first-order features).
The rationale for the second-order features is that since the first-order
features are selected manually based on Yahoo! experience, features
are highly correlated. Thus second-order features may capture aspects
not previously thought by feature designers. Since the number of second-order
features is large, we perform a correlation-based selection. First,
we compute the Pearson's correlation between each second-order feature
with the label, then choose those features whose absolute correlation
is beyond a threshold. For this particular data, we found the threshold
of $0.15$ is useful, although we did not perform an extensive search.
The number of selected second-order features is $14,188$.

\subsection{Results}

\begin{table}
\begin{centering}
\begin{tabular}{rccc|ccc}
 & \multicolumn{3}{c|}{First-order features} & \multicolumn{3}{c}{Second-order features}\tabularnewline
\cline{2-7} 
 & ERR & NG@1 & NG@5 & ERR & NG@1 & NG@5\tabularnewline
\hline
\hline 
\textbf{Rank Regress} & 0.4882 & 0.683 & 0.6672 & 0.4971 & 0.7021 & 0.6752\tabularnewline
\hline 
\textbf{RankNet} & 0.4919 & 0.6903 & 0.6698 & 0.5049 & 0.7183 & 0.6836\tabularnewline
\hline 
\textbf{Ranking SVM} & 0.4868 & 0.6797 & 0.6662 & 0.4970 & 0.7009 & 0.6733\tabularnewline
\hline 
\textbf{ListMLE} & 0.4955 & 0.6993 & 0.6705 & 0.5030 & 0.7172 & 0.6810\tabularnewline
\hline 
\textbf{PairTies-D} & 0.4941 & 0.6944 & 0.6725 & 0.5013 & 0.7131 & 0.6786\tabularnewline
\hline 
\textbf{PairTies-RK} & 0.4946 & 0.6970 & 0.6716 & 0.5030 & 0.7136 & 0.6793\tabularnewline
\hline
\hline 
\textbf{$\model$-FD} & 0.5038 & 0.7137 & 0.6762 & \textbf{0.5086} & \textbf{0.7272} & \textbf{0.6858}\tabularnewline
\hline 
\textbf{$\model$-Gibbs} & 0.5037 & 0.7105 & \textbf{0.6792} & 0.5040 & 0.7124 & 0.6706\tabularnewline
\hline 
\textbf{$\model$-MH} & \textbf{0.5045} & \textbf{0.7139} & 0.6790 & 0.5053 & 0.7122 & 0.6713\tabularnewline
\hline
\end{tabular}
\par\end{centering}

\caption{Performance measured in ERR and NDCG@T. PairTies-D and PairTies-RK
are the Davidson method and Rao-Kupper method for ties handling, respectively.
$\model$-FD is the $\model$ with full-decomposition, and $\model$-Gibbs/MH
is the $\model$ with Gibbs/Metropolis-Hasting sampling (see Section~\ref{sub:Learning-to-rank}
for a description). \label{tab:Performance}}

\end{table}

The results are reported in Table~\ref{tab:Performance}. The following
conclusions can be drawn. First, the use of second order features
improves the performance for nearly all the baseline methods. In our
algorithms, the second order features yield better performance for
PMOP-FD (incorporating the full decomposition).

Second, using either first or second order features, all our algorithms
outperform the baseline methods. For example, the $\model$-MH wins
over the best performing baseline, ListMLE, by $1.82\%$, using first-order
features. In our view, this is a significant improvement given the
scope of the dataset. We note that the difference in the top $20$
in the leaderboard of the Yahoo! challenge is just $1.56\%$. 

\begin{table}
\begin{centering}
\begin{tabular}{|c|c|}
\hline 
Pairwise models & $\model$/ListMLE\tabularnewline
\hline
\hline 
$\max\{\mathcal{O}(N^{2}),\mathcal{O}(NF)\}$ & $\mathcal{O}(NF)$\tabularnewline
\hline
\end{tabular}
\par\end{centering}

\caption{Learning complexity of models, where $F$ is the number of unique
features. For pairwise models, see Appendix~\ref{sec:Pairwise-Losses}
for the details. \label{tab:Learning-complexity}}

\end{table}

As for training time, the $\model$-FD is numerically the fastest
method. Theoretically, it has the linear complexity similar to ListMLE.
All other pairwise methods are quadratic in query size, and thus numerically
slower. The $\model$-Gibbs/MH is also linear in the query size, by
a constant factor that is determined by the number of iterations.
See Table~\ref{tab:Learning-complexity} for a summary.

\section{Conclusions}

Addressing the general problem of ranking with ties, we have proposed
a generative probabilistic model, with suitable parameterisation to
address the problem complexity. We present efficient algorithms for
learning and inference.We evaluate the proposed models on the problem
of learning to rank with the data from the currently held Yahoo! challenge.
demonstrating that the models are competitive against well-known rivals
designed specifically for the problem, both in predictive performance
and training time.

\bibliographystyle{plain}
\bibliography{../bibs/ME}

\appendix

\section{Computing $C$ \label{sec:Computing-constant}}

Let us calculate the constant $C$ in Eq (\ref{eq:decompose}). Let
use rewrite the equation for ease of comprehension

\[
\sum_{S\in2^{\remainset_{k}}}\frac{1}{\left|S\right|}\sum_{x\in S}\phi_{k}\left(x\right)=C\times\sum_{x\in R_{k}}\phi_{k}(x)\]
where $2^{R_{k}}$ is the power set with respect to the set $R_{k}$,
or the set of all non-empty subsets of $R_{k}$. Equivalently\[
C=\sum_{S\in2^{\remainset_{k}}}\frac{1}{\left|S\right|}\sum_{x\in S}\frac{\phi_{k}\left(x\right)}{\sum_{x\in R_{k}}\phi_{k}(x)}\]
If all objects are the same, then this can be simplified to\begin{eqnarray*}
C & = & \sum_{S\in2^{\remainset_{k}}}\frac{1}{\left|S\right|}\sum_{x\in S}\frac{1}{N_{k}}=\frac{1}{N_{k}}\sum_{S\in2^{\remainset_{k}}}1\\
 & = & \frac{2^{N_{k}}-1}{N_{k}}\end{eqnarray*}
where $N_{k}=|R_{k}|$. In the last equation, we have made use of
the fact that $\sum_{S\in2^{\remainset_{k}}}1$ is the number of all
possible non-empty subsets, or equivalently, the size of the power
set, which is known to be $2^{N_{k}}-1$. One way to derive this result
is the imagine a collection of $N_{k}$ variables, each has two states:
`\emph{selected}' and `\emph{not selected}', where `selected' means
the object belongs to a subset. Since there are $2^{N_{k}}$ such
configurations over all states, the number of non-empty subsets must
be $2^{N_{k}}-1$.

For arbitrary objects, let us examine the the probability that the
object $x$ belong to a subset of size $m$, which is $\frac{m}{N_{k}}$.
Recall from standard combinatorics that the number of $m$-element
subsets is the binomial coefficient $\left({N_{k}\atop m}\right)$,
where $1\le m\le N_{k}$, and . Thus the number of times an object
appears in any $m$-subset is $\left({N_{k}\atop m}\right)\frac{m}{N_{k}}$.
Taking into account that this number is weighted down by $m$ (i.e.
$|S|$ in Eq (\ref{eq:decompose})), the the contribution towards
$C$ is then $\left({N_{k}\atop m}\right)\frac{1}{N_{k}}$. Finally,
we can compute the constant $C$, which is the weighted number of
times an object belongs to any subset of any size, as follows\begin{eqnarray*}
C & = & \sum_{m=1}^{N_{k}}\left({N_{k}\atop m}\right)\frac{1}{N_{k}}=\frac{1}{N_{k}}\sum_{m=1}^{N_{k}}\left({N_{k}\atop m}\right)\\
 & = & \frac{2^{N_{k}}-1}{N_{k}}\end{eqnarray*}
We have made use of the known identity $\sum_{m=1}^{N_{k}}\left({N_{k}\atop m}\right)=2^{N_{k}}-1$.

\section{Pairwise Losses \label{sec:Pairwise-Losses}}

Let $\delta_{ij}(w)=\phi(x_{i},w)-\phi(x_{j},w)$, the pairwise losses
are\[
\mbox{loss}(x_{i}\succ x_{j};w)=\begin{cases}
\log(1+\exp(-\delta_{ij}(w))) & \,\,\mbox{ for\,\ logistic\,\ loss\,\ in\,\ RankNet}\\
\max\{0,1-\delta_{ij}(w)\} & \mbox{\,\,\ for\,\ hinge\,\ loss\,\ in\,\ Ranking\,\ SVM}\\
(1-\delta_{ij}(w))^{2} & \mbox{\,\,\ for\,\ quadratic\,\ loss\,\ in\,\ Pair\,\ Regress}\end{cases}\]
The overall loss is then\[
\mbox{Loss}=\sum_{i<j}\mbox{loss}(x_{i}\succ x_{j};w)\]
Taking derivative with respect to $w$ yields\begin{eqnarray*}
\frac{\partial\mbox{Loss}}{\partial w} & = & \sum_{i}\sum_{j|j<i}\frac{\partial\mbox{loss}(x_{i}\succ x_{j};w)}{\partial\delta_{ij}(w)}\left(\frac{\partial\phi(x_{i},w)}{\partial w}-\frac{\partial\phi(x_{j},w)}{\partial w}\right)\\
 & = & \sum_{i}\left(\sum_{j|j<i}\frac{\partial\mbox{loss}(x_{i}\succ x_{j};w)}{\partial\delta_{ij}(w)}\right)\frac{\partial\phi(x_{i},w)}{\partial w}-\sum_{j}\left(\sum_{i|i>j}\frac{\partial\mbox{loss}(x_{i}\succ x_{j};w)}{\partial\delta_{ij}(w)}\right)\frac{\partial\phi(x_{j},w)}{\partial w}\end{eqnarray*}
As it takes $N^{2}$ time to compute all the partial derivatives $\frac{\partial\mbox{loss}(x_{i}\succ x_{j};w)}{\partial\delta_{ij}(w)}$
for all $i,j$ where $j<i$, the overall gradient requires $N^{2}+NF$
time. Thus the complexity of the pairwise methods is $\mathcal{O}(\max\{N^{2},NF\})$.

\section{Learning the Paired Ties Models \label{sec:Learning-the-Paired-Ties}}

This section describes the details of learning the paired ties models
discussed in Section~\ref{sec:Discussion}.

\paragraph{Rao-Kupper method.}

Recall that the Rao-Kupper model defines the following probability
masses \begin{eqnarray*}
P(x_{i}\succ x_{j}|w) & = & \frac{\phi(x_{i},w)}{\phi(x_{i},w)+\theta\phi(x_{j},w)}\\
P(x_{i}\prec x_{j}|w) & = & \frac{\phi(x_{j},w)}{\theta\phi(x_{i},w)+\phi(x_{j},w)}\\
P(x_{i}\approx x_{j}|w) & = & \frac{(\theta^{2}-1)\phi(x_{i},w)\phi(x_{j},w)}{\left[\phi(x_{i},w)+\theta\phi(x_{j},w)\right]\left[\theta\phi(x_{i},w)+\phi(x_{j},w)\right]}\end{eqnarray*}
where $\theta\ge1$ is the ties factor and $w$ is the model parameter.
For ease of unconstrained optimisation, let $\theta=1+e^{\alpha}$
for $\alpha\in\mathbb{R}$. In learning, we want to estimate both
$\alpha$ and $w$. Let\begin{eqnarray*}
P_{i} & = & \frac{\phi(x_{i},w)}{\phi(x_{i},w)+(1+e^{\alpha})\phi(x_{j},w)}\\
P_{j}^{*} & = & \frac{\phi(x_{j},w)}{\phi(x_{i},w)+(1+e^{\alpha})\phi(x_{j},w)}\\
P_{i}^{*} & = & \frac{\phi(x_{i},w)}{(1+e^{\alpha})\phi(x_{i},w)+\phi(x_{j},w)}\\
P_{j} & = & \frac{\phi(x_{j},w)}{(1+e^{\alpha})\phi(x_{i},w)+\phi(x_{j},w)}\end{eqnarray*}

Taking partial derivatives of the log-likelihood gives\begin{eqnarray*}
\frac{\partial\log P(x_{i}\succ x_{j}|w)}{\partial w} & = & (1-P_{i})\frac{\partial\log\phi(x_{i},w)}{\partial w}-(1+e^{\alpha})P_{j}\frac{\partial\log\phi(x_{j},w)}{\partial w}\\
\frac{\partial\log P(x_{i}\succ x_{j}|w)}{\partial\alpha} & = & -P_{j}e^{\alpha}\\
\frac{\partial\log P(x_{i}\approx x_{j}|w)}{\partial w} & = & (1-P_{i}-(1+e^{\alpha})P_{i}^{*})\frac{\partial\log\phi(x_{i},w)}{\partial w}+(1-P_{j}-(1+e^{\alpha})P_{j}^{*})\frac{\partial\log\phi(x_{j},w)}{\partial w}\\
\frac{\partial\log P(x_{i}\approx x_{j}|w)}{\partial\alpha} & = & \left(\frac{2(1+e^{\alpha})}{(1+e^{\alpha})^{2}-1}-P_{i}^{*}-P_{j}^{*}\right)e^{\alpha}\end{eqnarray*}

\paragraph{Davidson method.}

Recall that in the Davidson method the probability masses are defined
as\begin{eqnarray*}
P(x_{i}\succ x_{j}|w) & = & \frac{\phi(x_{i},w)}{\phi(x_{i},w)+\phi(x_{j},w)+\nu\sqrt{\phi(x_{i},w)\phi(x_{j},w)}}\\
P(x_{i}\prec x_{j}|w) & = & \frac{\phi(x_{j},w)}{\phi(x_{i},w)+\phi(x_{j},w)+\nu\sqrt{\phi(x_{i},w)\phi(x_{j},w)}}\\
P(x_{i}\approx x_{j}|w) & = & \frac{\nu\sqrt{\phi(x_{i},w)\phi(x_{j},w)}}{\phi(x_{i},w)+\phi(x_{j},w)+\nu\sqrt{\phi(x_{i},w)\phi(x_{j},w)}}\end{eqnarray*}
where $\nu\ge0$. Again, for simplicity of unconstrained optimisation,
let $\nu=e^{\beta}$ for $\beta\in\mathbb{R}$. Let\begin{eqnarray*}
P_{i} & = & \frac{\phi(x_{i},w)}{\phi(x_{i},w)+\phi(x_{j},w)+e^{\beta}\sqrt{\phi(x_{i},w)\phi(x_{j},w)}}\\
P_{j} & = & \frac{\phi(x_{j},w)}{\phi(x_{i},w)+\phi(x_{j},w)+e^{\beta}\sqrt{\phi(x_{i},w)\phi(x_{j},w)}}\\
P_{ij} & = & \frac{e^{\beta}\sqrt{\phi(x_{i},w)\phi(x_{j},w)}}{\phi(x_{i},w)+\phi(x_{j},w)+e^{\beta}\sqrt{\phi(x_{i},w)\phi(x_{j},w)}}\end{eqnarray*}
Taking derivatives of the log-likelihood gives\begin{eqnarray*}
\frac{\partial\log P(x_{i}\succ x_{j}|w)}{\partial w} & = & (1-P_{i}-0.5P_{ij})\frac{\partial\log\phi(x_{i},w)}{\partial w}-(P_{i}+0.5P_{ij})\frac{\partial\log\phi(x_{j},w)}{\partial w}\\
\frac{\partial\log P(x_{i}\succ x_{j}|w)}{\partial\beta} & = & -P_{ij}\\
\frac{\partial\log P(x_{i}\approx x_{j}|w)}{\partial w} & = & (0.5-P_{i}-0.5P_{ij})\frac{\partial\log\phi(x_{i},w)}{\partial w}+(0.5-P_{j}-0.5P_{ij})\frac{\partial\log\phi(x_{j},w)}{\partial w}\\
\frac{\partial\log P(x_{i}\approx x_{j}|w)}{\partial\beta} & = & 1-P_{ij}\end{eqnarray*}

\end{document}